# Random Network Behaviour of Protein Structures

Brinda K.V.[1], Saraswathi Vishveshwara[2] and Smitha Vishveshwara[3]

[1] Department of Chemistry and Biochemistry, University of Texas at Austin, Austin, TX 78712, USA.

[2] Molecular Biophysics Unit, Indian Institute of Science, Bangalore, 560012, India.

[3]1110 W. Green Street, Dept. of Physics, University of Illinois at Urbana-Champaign, Urbana IL 61801, USA.

*Corresponding author: Smitha Vishveshwara, E-mail: smivish@illinois.edu*
*Supplementary Material can be found in the journal version in Molecular BioSystems.*

**Abstract:**
**Geometric and structural constraints greatly restrict the selection of folds adapted by protein backbones, and yet, folded proteins show an astounding diversity in functionality. For structure to have any bearing on function, it is thus imperative that, apart from the protein backbone, other tunable degrees of freedom be accountable. Here, we focus on side-chain interactions, which non-covalently link amino acids in folded proteins to form a network structure. At a coarse-grained level, we show that the network conforms remarkably well to realizations of random graphs and displays associated percolation behavior. Thus, within the rigid framework of the protein backbone that restricts the structure space, the side-chain interactions exhibit an element of randomness, which account for the functional flexibility and diversity shown by proteins. However, at a finer level, the network exhibits deviations from these random graphs which, as we demonstrate for a few specific examples, reflect the intrinsic uniqueness in the structure and stability, and perhaps specificity in the functioning of biological proteins.**

**Key words:** protein structure network, non-covalent connections, probabilistic distribution, percolation transition, giant cluster

## Introduction:

A protein is a hetero-polymer composed of a sequence of amino acids, which, among billions of possibilities for putative configurations, stunningly assumes a unique structure, whose precise functions govern life's processes (1). It is well known that proteins respect severe constraints imposed by folding entropy (2) resulting in a limited menu of protein folds (3). The backbone of the polypeptide chain endows the protein a skeletal structure composed of optimally packed (4), immutable folds (5, 6), which are resilient to local variations and mutations (7, 8). Moreover, the underlying structure of amino acid linkages formed via non-covalent side-chain interactions is also known to be crucial for the stability and uniqueness of protein structure. While the backbone accounts for robustness of structure, its regular packing alone explains neither the diversity of sequences for a given fold, nor functional specificity and diversity of proteins. However, the role of side-chain linkages in this regard has received much less attention. In the present work, by analyzing a large dataset of protein structures, we find that the three-dimensional network (9-12) formed by these amino acid side chain links exhibits features of randomness (13). Although randomness has been established in amino acid sequences (7, 8, 14), it has only sparsely been investigated in the context of interactions in spatial structure in proteins (15,16,17). For example, Bryngelson and Wolynes have introduced earlier a random energy model for understanding the nature of the folding energy landscape of proteins (18,19). This phenomenological model has established the concepts of ruggedness and smoothness in the folding energy landscape and has also provided a way for understanding the kinetics of protein folding. The present study, which is based on experimentally determined protein structures, shows that the non-covalent interactions in their native state structures have elements of randomness as seen by the percolation behaviour of the amino acid networks in protein native structures. And the results underscore the presence of order, reflected in the presence of a rigid backbone, coexisting with disorder, reflected in the random percolation-like behaviour of the side-chains, in protein structures. We suggest that the interplay between order and disorder yields stability, on the one hand, and sensitivity towards changes such as in cellular environment and ligand binding, on the other, to protein structures. Further, this random behavior, or more precisely, a probabilistic distribution for the formation of links within the protein structure, provides an extensive parameter space to host variations while conforming to structural, chemical and biological constraints. Hence, we believe that the side chain linkages, within the framework of the backbone architecture, offer the degrees of freedom required to host a tremendous range of specific structures which may be crucial in accounting for the marvelous diversity observed in Nature's functioning proteins. Furthermore, the deviations to the random network model also account for the specific and unique functioning of the diverse range of proteins.

**Results and Discussion:**

The connectivity of the amino acid networks within the protein native structure depends strongly on the manner in which the widely varying side-chain interaction strength is quantified. We quantify the interaction strength $I_{ij}$, in globular proteins based on the number of atom-atom contacts made by the two interacting amino acids 'i' and 'j'. The amino acids, considered as nodes in a graph, are connected by an edge if $I_{ij}$ is greater than a specified minimum interaction strength $I_{min}$ (Fig. 1a, 1b, ref. 11, 12 and supplement section). We have earlier analyzed these protein structure graphs for more than 200 representative globular proteins (12) and we use a size-dependent subset of these proteins in the present study. In what follows, we show that protein structure networks thus formed by the side-chains are primarily random graphs wherein amino acid nodes are linked to one another with a probability 'p' which depends on the specified interaction strength $I_{min}$. We also discuss the implications of the deviations from the random model and the biological significance.

**(i) Degree distribution shows Erdos-Renyi like random model**

A signature feature identifying properties of a network is the degree distribution, the degree being the number of links connected to a node. To study the degree distribution of side chain networks, we selected sets of globular protein structures from the RCSB protein data bank (20) of varying sizes ranging from N = 100 to N=1000. Following the schemes of ref. (12), we plotted the degree distributions of the amino acids for each set of proteins as a function of $I_{min}$ (Fig. 2). We find that the degree distribution and other network features, such as the links/node ratio, the size of the largest cluster, and the presence of highly connected nodes called "hubs", follow the same qualitative behavior for a wide range of proteins (also observed for the 232 proteins studied in ref 12). The degree distributions do not fit power-law, exponential, or Gaussian distributions or distributions characteristic of regularly packed structures. It must be noted that the incompatibility with the power-law distribution immediately eliminates the scale-free behaviour attributed to several real world networks such as the social networks (21), World Wide Web and metabolic networks (22). However, the degree distributions fit well to the Poisson distribution, characteristic of random processes, given by

$$n(k) = C \lambda^k e^{-\lambda} /k! , \qquad (1)$$

where n is the number of nodes with 'k' edges and 'C' and 'λ' are fitting parameters. The Poisson form given by equation (1) in fact exactly corresponds to the degree distribution predicted by one of the simplest paradigm random network models - the Erdõs-Rényi (ER) model (23) which consists of a set of N nodes, each having the same probability p of forming an edge with another. In the ER model, the parameters C and λ represent N and pN, respectively. Hence, for a given size N, the only fitting parameter required to make a correspondence between the protein data for a particular $I_{min}$ and the ER model, is the probability of connection, 'p'. The degree distribution of the protein data for a set of $I_{min}$ and the best fit ER Poisson curves are shown in Fig. 2. The behaviour of the number of nodes without edges (k = 0) is also shown and to first order, is remarkably consistent with expectations for the ER model. Thus, a given value of $I_{min}$ in proteins directly corresponds to the probability 'p' of forming an edge in a random graph; the larger the $I_{min}$, the lower the probability that two amino acids are linked (specific mapping is given in Fig. S1 in supplement section).

**(ii) Giant cluster shows percolation-like transition**

A hallmark of a broad class of random networks is the presence of a transition point at which a giant cluster percolates the system; here, we investigate such a percolation transition for the largest connected cluster as a function of $I_{min}$. Strikingly, as shown in Fig. 3a, we find that the size of this largest cluster for protein structures follows a sigmoidal profile for all protein sizes and indeed a transition occurs within a narrow range of $I_{min}$ indicated by a sudden drop in the largest cluster size (the same is observed for the 232 proteins studied in ref 12). In the ER model (23,24), the size of the giant cluster $N_{Gc}$ at this transition point and the critical probability $p_c$ for the transition vary as

$$N_{Gc} \approx N^{2/3}, \ p_c \approx 1/N. \qquad (2)$$

The ER criterion for the giant cluster behavior provides a good quantitative estimate of the transition point in proteins; the sharp drop in $N_{Gc}$ in Fig. 3a occurs roughly when the size of the giant cluster takes on the value $N^{2/3}$, where N is the number of amino acid residues in the protein. Furthermore, upon using the mapping derived above between $I_{min}$ and probability p, we find that as shown in Fig. 3b, the critical probability $p_c$ at which the transition occurs is surprisingly close to the 1/N estimate given by the ER model. The presence of the transition gives a clear measure of how connected and tightly bound the amino acids are within the protein, which in turn is related to the packing and stability of the protein. Indeed, previous studies have shown that proteins are packed in a liquid-like random fashion close to the percolation transition (25); the percolation behaviour of the protein network studied here ought to be intimately related to such packing properties and a rigorous mapping between the two pictures is in order.

**(iii) Deviations from random model**

While the broad features of the protein structure networks, like the percolation of the giant cluster and the Poisson fit of the degree distribution, conform to those of the ER model, deviations are abound. Moreover, the associated percolation fails to show universal behaviour associated with phase transitions; we performed a finite-size scaling analysis which did not show good data collapse (24). One obvious cause for these deviations is the presence of an underlying peptide-linked backbone in protein structures; amino-acids lying sequentially on the polypeptide chain are a priori linked. Numerical simulations of a modified ER model, which accounted for the presence of the backbone, matched some of the deviations observed in the Poisson fit of Fig. 2, in particular, the deviation of the peak value of the degree distribution (Fig. S2 in supplement section). This shows that incorporating small but essential features of proteins into the random graph models, in this case, constraints posed by the backbone, captures trends displayed by the protein structure networks.

Another reason for deviation in degree distribution is the topology of the protein, which is approximately spherical and hence the peripheral nodes selectively have less number of edges resulting in higher number of orphans (at low $I_{min}$) in proteins. It would also be interesting to study higher order percolation behaviour such as clique percolation and examine if deviations from the random network are seen at such higher order levels. Yet another important cause for the deviations is that proteins, being hetero-polymers, have non-uniform probabilistic interactions between amino acids. Residue-specific interaction biases have been studied for a large protein dataset (26-27) and can be incorporated in a random model by treating the probability 'p' not as homogeneous but as amino-acid specific.

These deviations from random models, while potentially stemming from effects at the microscopic level, reflect macroscopic biological features, as we now demonstrate in the following examples. Quantitatively different giant cluster profiles and percolation features are seen for two sets of proteins of almost identical size, structure and function, but differing in biological properties. We discuss the following example of thermal stability and the percolation transition in some detail.

**(a) Thermal stability of proteins**

Let us consider the example of enhanced stability of thermophilic proteins over mesophilic ones. A thermophilic protein is stable at higher temperatures, whereas its mesophilic counterpart is stable only at ambient temperatures. Hence, proteins of similar size and structure, performing the same function, can differ in their thermal stability. We obtained a set of 14 thermophilic proteins and their mesophilic counterparts and studied their protein structure networks in terms of degree distributions and largest cluster sizes and other network parameters. The preliminary results of this study have been published earlier (12). Here, we delve a little further to identify the critical $I_{min}$ ($I_c$) and critical 'p' values ($P_c$) for each of these proteins from their largest cluster plots, which are typically sigmoidal in nature with a critical transition point. According to the ER random model, this critical transition point is identified as the point where the largest cluster size is approximately equal to $N^{2/3}$, where N is the size of the protein in terms of number of residues (22,24). Table 1 provides the $I_c$ and $P_c$ values identified for the chosen thermophilic proteins and their mesophilic counterparts.

We find that the sizes of thermophilic and corresponding mesophilic proteins are comparable although not the same. Further, in general, we observe that when the largest cluster size = $N^{2/3}$, $p=P_c=1/N$., the thermophilic proteins have higher $I_c$ and lower $P_c$ than the corresponding mesophilic proteins [Note: I and P are inversely correlated].As an example, we show the giant cluster profiles for the thermophilic and the mesophilic carboxy-peptidase in Fig. 4. It can be seen that the giant cluster profiles of the two proteins are qualitatively different. Further, the highly stable thermophilic protein has a lower $p_c$ than its mesophilic counterpart. The increased stability of the thermophilic protein is beautifully explained by stronger interactions (lower $P_c$ and corresponding higher $I_{min}$) involved in the formation of the largest cluster at the transition point.

The deviations occur in case of Phosphofructo Kinase and Glyceraldehyde-3-Phosphate dehydrogenase, where the mesophilic protein has a higher $I_c$ and lower $P_c$ than the thermophilic protein (Table 1). And in Phosphoglycerate kinase, both the thermophilic and mesophilic proteins have the same $I_c$ and $P_c$ values. From a percolation point of view, a lower $P_c$ value, indicates a network that is more stable (22). Similarly higher $I_c$ and lower $P_c$ observed in thermophilic proteins, can indicate a more stable protein structure network, leading to the higher temperature stability of the thermophilic proteins. If the protein structure networks of these proteins completely adhered to the ER random model, then their largest cluster profiles and specifically, their $p_c$ ($\approx 1/N$) would be almost identical since they are similar in size and structure. However, we find that the giant cluster profiles and the $p_c$, although similar, are not nearly identical and therefore, we believe that the biological demand of additional stability is achieved as a deviation from the random behavior. The two cases of Phosphofructokinase and Glyceraldehyde-3-Phosphate dehydrogenase are likely exceptions, where the additional thermal stability is not accounted by deviations from the random network behavior.

It is commonly believed that additional thermal stability can be achieved by making stronger or more interactions between the amino acid residues in protein native structures. These interactions are generally hydrogen bonds, salt bridges or di-sulfide bridges as have been shown in earlier studies (28-29). We have listed the number of such interactions in the pairs of structures (given in Table 1) in the supplementary section (Table S2). We find that, the number of di-sulphide bridges is very small and not seen in most proteins. Further, we also observe that, in many cases the thermophilic protein has more number of salt bridges and sidechain-sidechain and sidechain-mainchain hydrogen bonds and in general, more number of interactions as expected. However, this is not true in cases like triose phosphate isomerase, neutral protease and lactate dehydrogenase, where our network representation and the deviations from the ER model are able to provide a better picture of the additional stability (Table 1). Further, the cases in which our network representation does not account for the additional stability like Phosphofructo Kinase and Glyceraldehyde-3-Phosphate dehydrogenase, we find that the number of interactions given in Table S1 also falls short. Hence, the network representation provides additional insights into thermal stability of proteins not accounted by the number of interactions alone provided in Table S2. This is probably because the network representation provides a global picture of the amino acid interactions (including strong and weak) in the protein structure rather than a sum of pair-wise interactions of hydrogen bonds or salt-bridges. Our network model may further be refined above the ER model, if the connections are weighed on the basis of energy terms such as hydrogen bonds, salt bridges and di-sulphide bridges. This aspect is currently under investigation in our laboratory (Vijaybaskar & Vishveshwara, work in progress).

**(b) Ligand-binding in proteins**

Another deviation to the ER random model that has strong biological implications is the effect of binding of small molecules or ligands to proteins. Most proteins bind to ligands during the course of their life in the cell and many times it is a functional requirement for the protein to bind to a ligand. We show the giant cluster profiles for fully liganded Glutaminyl t-RNA synthetase (bound to the ligands glutaminyl-tRNA, ATP analog and Glutamine) and the corresponding ligand free protein in Fig 5. The overall giant cluster profiles of the two proteins are similar, as can be expected, since they are the same protein to begin with. However, subtle differences in the plot can be observed, especially in the $p_c$ values, where the ligand-bound protein shows lower $p_c$ corresponding to higher stability as compared to ligand-free protein. Again, proteins of the same size, structure and function show subtle variations in the percolation properties indicating that they are inherently able to adapt to biological needs by exhibiting specificity as well as sensitivity to small changes like the binding of a ligand.

As seen in these examples of thermal stability and ligand binding, deviations from random behavior are influenced by strong biophysical demands. One can expect such deviations to occur at other instances of biological demand such as stabilization of multiple conformations in proteins, specificity of enzyme catalysis, and resistance of proteins to chemical and biological factors in the cell. In fact, we find that although all globular proteins qualitatively show similar percolation-like random behavior, rarely do proteins of the same size exhibit the same percolation properties. On the one hand, the percolation-like randomness confers the sensitivity and adaptability to the protein structures within the rigid framework of the protein backbone. However, on the other hand, specificity and stability required for proper functioning of the protein are achieved as a consequence of deviations from the random behavior.

**Conclusions:**

Our studies clearly show that, to a large extent, the side-chain linked protein networks display random behavior, suggesting that they are primarily the most likely configurations that satisfy the constraints of connectivity. Equally important are the deviations from the random model brought about by chemical and biological factors, and the dynamics of protein folding. The interplay between the randomness and orderliness, evident from the percolation-like random behavior and the deviations within, account for the diversity and sensitivity, specificity and uniqueness, observed in protein structure, stability and function. A full development of this data-based random graph picture would provide insight into several significant problems such as the optimal conditions for protein stability and functioning, the connection between protein structure and function, and the mapping of an expansive sequence space to a restricted structure space.

**Methods:**

**(a) Construction of protein structure networks (PSN):**

The method of constructing protein structure networks (including the definition of $I_{min}$) and their analyses using degree distributions and largest cluster plots have been described in detail in references 11 and 12 and a summary is provided in the supplement section. We construct protein structure graphs by considering each amino acid in the protein structure as a node and the non-covalent spatial interactions between the amino acid side-chains constitute the edges in the graph. The strength of the interaction between the amino acid side-chains are evaluated based on the number of side-chain atoms that come within interacting distance of each other and a minimum interaction strength cutoff ($I_{min}$) is a variable used in deciding the connected edges in the protein structure graph (refer supplement section for details).

**b) Poisson fitting for protein graphs:**

The nature of the degree distribution in the protein structure graphs (Fig. 2) is reminiscent of the Poisson distribution of the well-known Erdõs-Rényi model of random graphs given by

$$y(k)=(pN)^k e^{-pN}/k!$$

where y(k) is the probability of finding a node having 'k' edges.

The poisson-like features of the degree distribution and the percolation-type features of the largest protein cluster (Fig. 3) compel us to map the protein network to the ER model random graph, as a first order means of investigating randomness in proteins. We thus fit the protein degree distribution to the Poisson form described above, with $I_{min}$ playing the role of the probability of forming an edge ‘p’ and N the number of amino acids within a protein. As seen from the above expression, the Poisson distribution is dependent upon the size of the network i.e., the number of nodes comprising the network, N. In order to fit our data to the Poisson curve, we chose a subset of 20 proteins (from the original representative set of 232 proteins used in ref 12) each of approximate sizes 200, 300, 400 and 1000 amino acid residues and construct the protein structure graphs as mentioned above. To establish the Poisson distribution for protein graphs, we establish a correlation between $I_{min}$ and the probability of forming and edge ‘p’ for each size bin. This was done by fitting the data from the protein graphs into the Poisson equation (using Matlab). Fig. S1 (supplementary information) provides the correlation of $I_{min}$ to p in protein structure graphs of varying sizes. We find that ‘p’ varies with the size of the protein for a given $I_{min}$ as expected for a Poisson distribution. Further, it is clear that $I_{min}$ and ‘p’ are inversely correlated, with higher $I_{min}$ corresponding to lower ‘p’ and vice-versa, for all protein sizes.

**c) Simulations of modified ER model**

Towards a better understanding of the correlation of the protein structure graphs with the ER model, we numerically generated random graphs and carried out studies similar to the protein structure graphs on them. To mimic our protein analysis, the connectivities with immediate neighbors (backbone) were deliberately ignored (so as to mimic the non-covalent side-chain network in proteins) and the condition of symmetry was imposed. We generated 20 random graphs (using Matlab) each of sizes 200, 300, 400 and 1000 nodes with varying probabilities of edge formation ‘p’ ($0.0005 < p < 0.04$). These probabilities for generating random graphs were chosen so as to reproduce the probability ranges characterizing the protein structure graphs. Detailed comparisons of the degree distribution and the percolation behaviour of the randomly generated graphs with the corresponding protein structure graphs are made in the supplementary information.

**Acknowledgements:** This work was supported by the National Science Foundation Grant No. DMR 06-44022 CAR and the Mathematical Biology project funded by the Department of Science and Technology (DST), India for computational facilities. We thank Dhruba Deb and Vijayabaskar MS for help in manuscript preparation.

**Table 1: Results from thermophile/mesophile analysis**

| Protein | Thermophile | Size | $I_c$[1] | $P_c$[2] | Mesophile | Size | $I_c$ | $P_c$ |
|---|---|---|---|---|---|---|---|---|
| Tata box binding protein | 1PCZ | 183 | 4.50 | 0.00528 | 1VOK | 192 | 4.25 | 0.00585 |
| Adenylate Kinase | 1ZIP | 217 | 4.25 | 0.00585 | 1AK2 | 220 | 4.00 | 0.00652 |
| Subtilisin | 1THM | 279 | 4.00 | 0.00469 | 1ST3 | 269 | 3.75 | 0.00512 |
| Carboxy peptidase | 1OBR | 323 | 6.00 | 0.00266 | 2CTC | 307 | 4.25 | 0.00417 |
| Neutral Protease | 1THL | 316 | 5.00 | 0.00343 | 1NPC | 317 | 4.50 | 0.00383 |
| Phospho fructo kinase | 3PFK | 319 | 4.00 | 0.00469 | 2PFK | 300 | 4.25 | 0.00417 |
| Lactate dehydrogenase | 1LDN | 316 | 4.50 | 0.00383 | 1LDM | 329 | 4.25 | 0.00417 |
| Glyceraldehyde-3-Phosphate-dehydrogenase | 1GD1 | 331 | 4.50 | 0.00383 | 1GAD | 327 | 5.00 | 0.00343 |
| Phospho glycerate Kinase | 1PHP | 394 | 3.75 | 0.00373 | 3PGK | 415 | 3.75 | 0.00373 |
| Reductase | 1EBD | 455 | 3.75 | 0.00373 | 1LVL | 458 | 3.00 | 0.00464 |
| TFIIB transcription factor | 1AIS | 193 | 4.75 | 0.00511 | 1VOL | 204 | 4.50 | 0.00528 |
| Xylanase | 1YNA | 193 | 5.75 | 0.00387 | 1XYN | 178 | 5.00 | 0.00469 |
| Triose phosphate isomerase | 1BTM | 251 | 5.00 | 0.00469 | 7TIM | 247 | 4.50 | 0.00528 |
| Signal recognition particle | 1FFH | 287 | 5.00 | 0.00343 | 1FTS | 295 | 4.25 | 0.00417 |

[1]: $I_c$ is the critical $I_{min}$
[2]: $P_c$ is the critical probability of forming an edge

# FIGURES

**Figure1:**

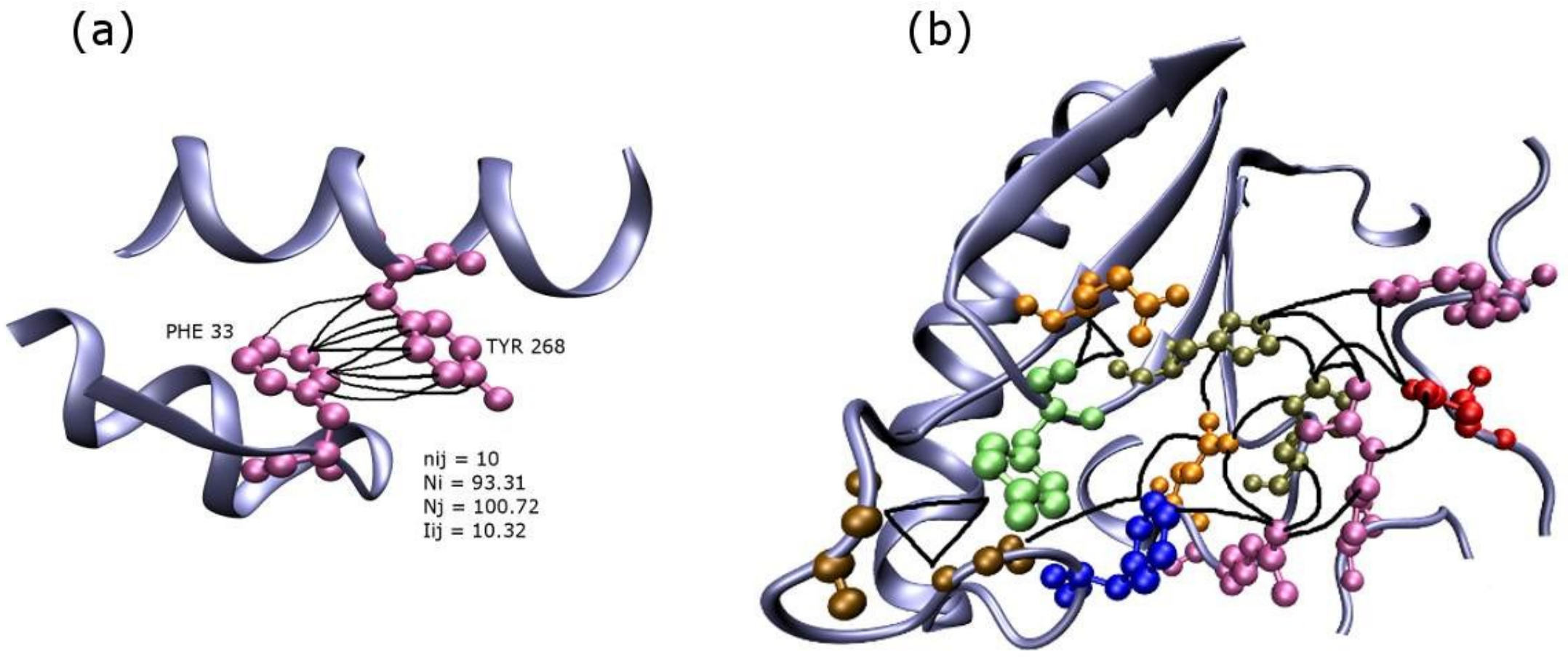

Non-covalent interactions in proteins: **(a)** Evaluation of the strength of non-covalent interaction as in Ref 11. The interaction between the two residues Phenylalanine 33 (i) and Tyrosine 268(j) in L-arabinose binding protein (Pdb code: 8ABP) is shown, in which the residues i and j are separated in sequence ((i-j)>2). The protein backbone and the interacting residues are shown in blue and pink respectively. The interacting atom pairs (distance less than 4.5 Å) are connected with thin black lines. Here $\mathbf{n_{ij}}$ is the number of distinct atom pairs between the residues i and j, while $N_i$ and $N_j$ are the normalization of the two residue types Phenylalanine and Tyrosine. $I_{ij}$ is the strength of interaction between the residues i and j, which is evaluated as $[n_{ij}/\mathrm{sqrt}(N_i \times N_j)] \times 100$. A protein structure graph of a desired interaction cutoff, $I_{min}$, is constructed by considering each of the residues as nodes and making an edge between any two residues with $I_{ij} > I_{min.}$ **(b)** Example of a protein structure graph: A networked cluster (with 12 residues obtained from the protein carboxypeptidase (2CTC) at $I_{min}$=6%) in the protein structure is depicted. The protein backbone is shown in blue cartoon and the amino acid residues contributing to the cluster are shown in different colored ball and stick representation. The interacting amino acid pairs are linked with black lines.

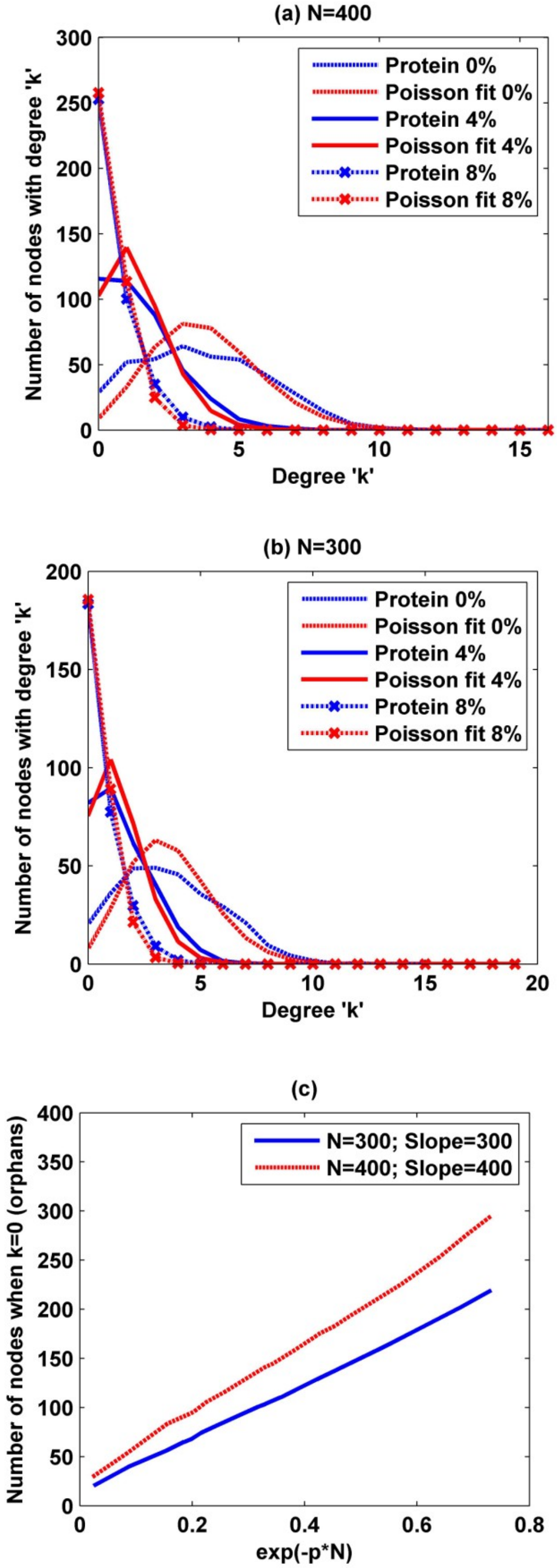
(a) N=400
Number of nodes with degree 'k'
Degree 'k'
Protein 0%
Poisson fit 0%
Protein 4%
Poisson fit 4%
Protein 8%
Poisson fit 8%
(b) N=300
Number of nodes with degree 'k'
Degree 'k'
Protein 0%
Poisson fit 0%
Protein 4%
Poisson fit 4%
Protein 8%
Poisson fit 8%
(c)
Number of nodes when k=0 (orphans)
exp(-p*N)
N=300; Slope=300
N=400; Slope=400

**Figure 2**: The degree (k, number of edges connected to a node) distribution plots (averaged) for proteins of size (a) N=400 and (b) N=300. Blue and red lines represent the data from proteins and their corresponding Poisson fits (obtained from Matlab) respectively at three different $I_{min}$s, 0% (dashed), 4% (line) and 8% (dotted). The Poisson distribution for the number of nodes 'n' having 'k' edges is given by $n(k) = C \lambda^k e^{-\lambda} /k!$, where C = N and are $\lambda = pN$ for ER networks. These plots show that the shape of the degree distribution is dependent on the $I_{min}$ irrespective of the protein size. Further, the protein graphs fit the ER distribution and each $I_{min}$ can be identified with a probability 'p', which is the only fitting parameter. (c) Plot of number of nodes without edges (orphans) in protein structure graphs versus exp(-pN) when N=300 (blue line) and N=400 (red dashed). The straight line obtained in this plot indicates that $n(0) \approx Ne^{-pN}$, thus, to first order, consistent with behavior of ER random graphs given by equation (1).

**Figure 3:**

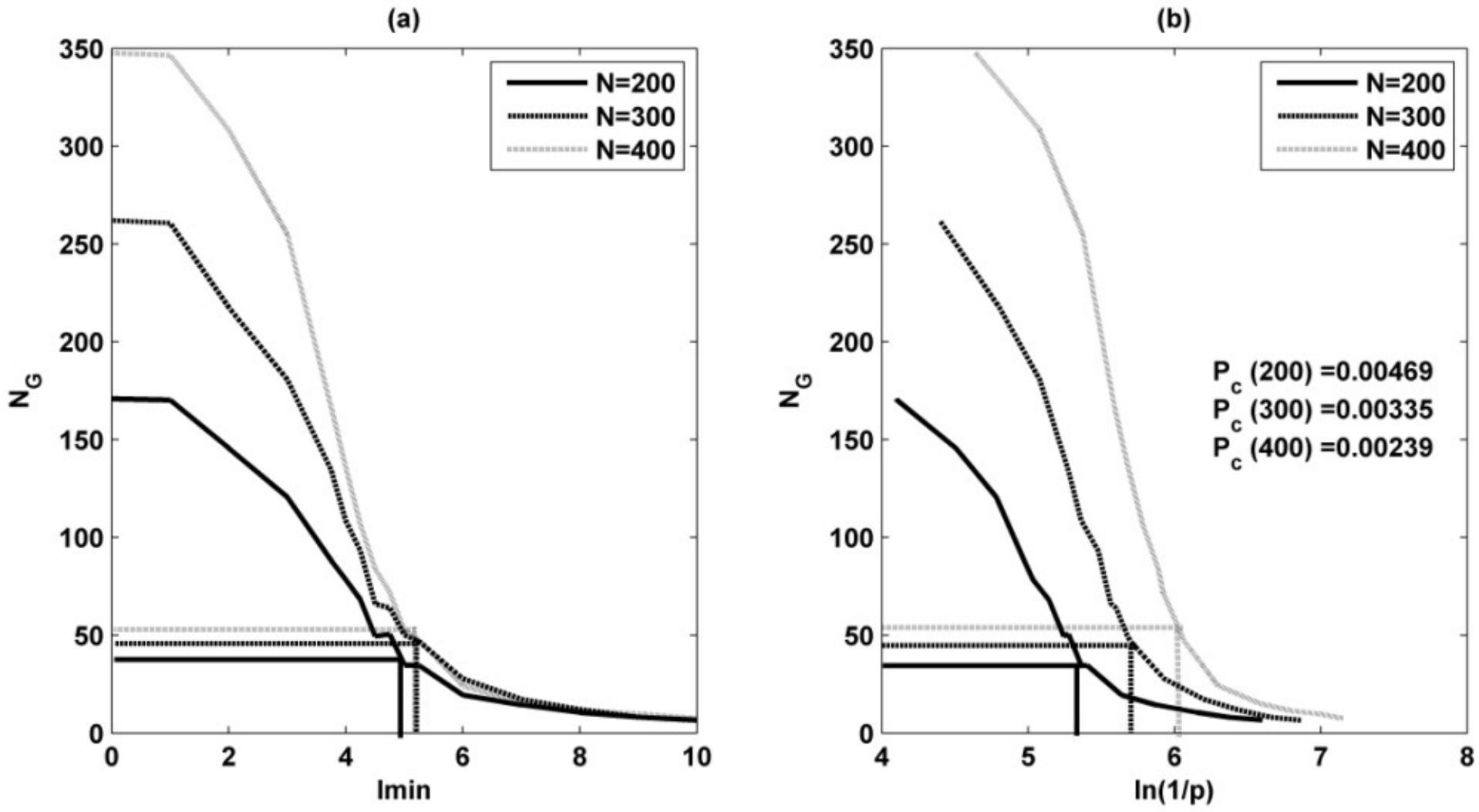

Largest cluster size plots. (a) Size of the largest cluster (averaged) in the protein structure graphs ($N_G$) versus $I_{min}$ for three different protein sizes, 200, 300 and 400. (b) Same as (a) but $N_G$ is plotted versus ln(1/p) after 'p' is correlated to $I_{min}$ for the different protein sizes by comparing with the Erdos-Renyi model. Plots show that $N_G$ undergoes a transition as a function of $I_{min}$ or ln(1/p) for all protein sizes, indicated by a sharp rise of cluster size below a certain $I_{min}$ at which a giant cluster permeates the system. The critical transition point, $P_c$ is identified as the point on the curve where $N_G \approx N^{2/3}$ (based on the criterion for the ER model) and is indicated for all three protein sizes (A similar plot for random graphs is given in the supplement section, Fig. S3). Moreover, similar to ER model predictions, $P_c \approx 1/N$ for these protein graphs. The $I_{min}$ corresponding to $P_c$ is shown in Figure 3a.

**Figure 4:**

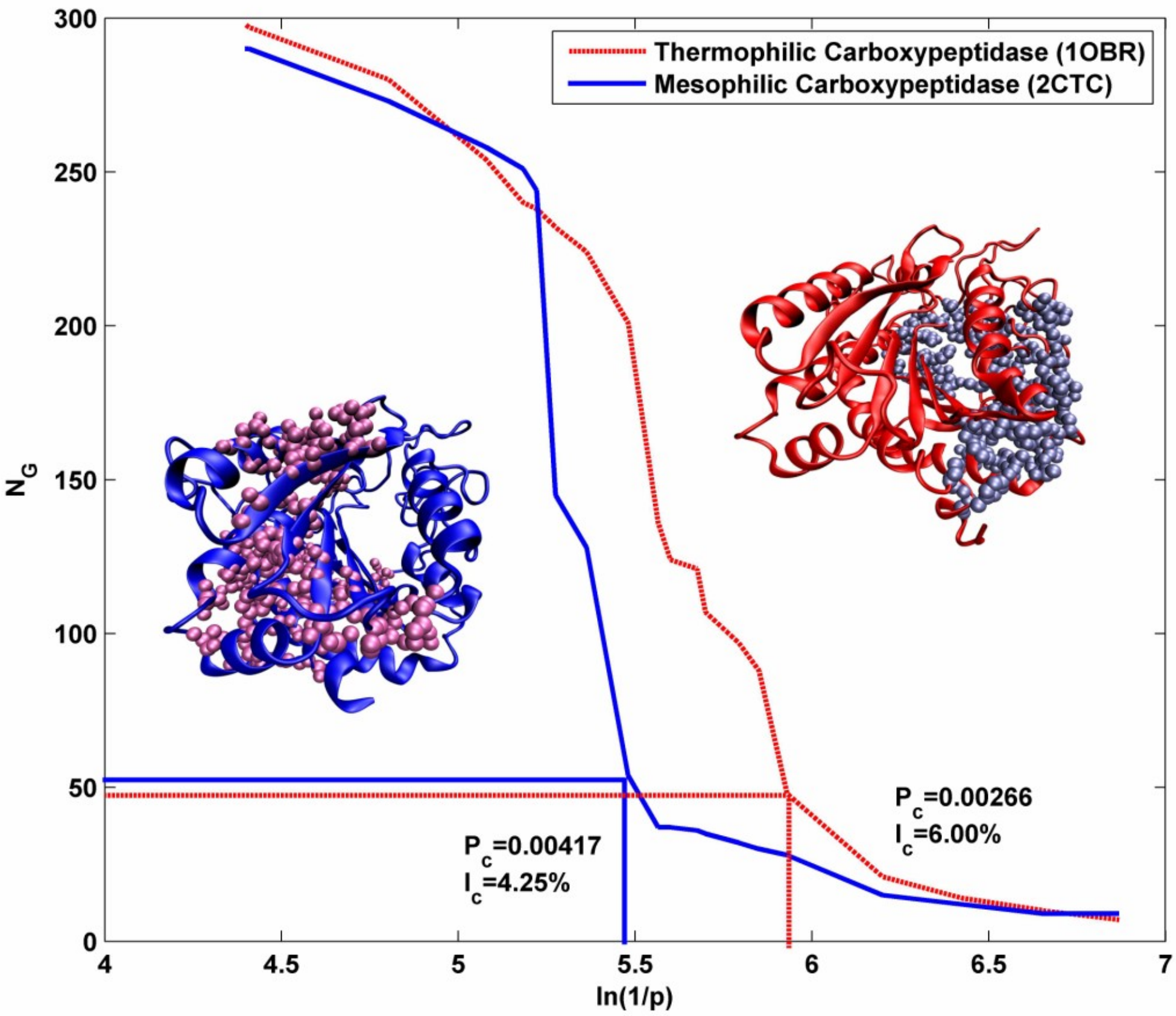

The largest cluster ($N_G$) plot versus ln(1/p) for the thermophilic protein (Carboxypeptidase, PDB code: 1OBR (red dashed)) and the corresponding mesophilic protein (Pdb code: 2CTC ( in blue line)). The critical 'p' and '$I_{min}$' values ($P_c$ and $I_c$ respectively) for both proteins are indicated in the plot. It is evident that the thermophilic protein has a higher $I_c$ and lower $P_c$ than the corresponding mesophilic protein, although both proteins are of fairly similar size (N=326 for 1OBR and N=307 for 2CTC). The cartoon representation of both protein structures (1OBR in red and 2CTC in blue) along with the residues in the largest cluster obtained at their respective $P_c$ (colored van der Waal's spheres) are also shown and these are found to be qualitatively different and occupy different regions of the protein structure. While the ER model forms a good basis for characterizing random behavior in protein networks, the example here of the significantly different behavior of two actual proteins of very similar size and composition shows that studying deviations from the ER model would be valuable for understanding the role of biological factors and features that determine the stability and functioning of proteins.

**Figure 5:**

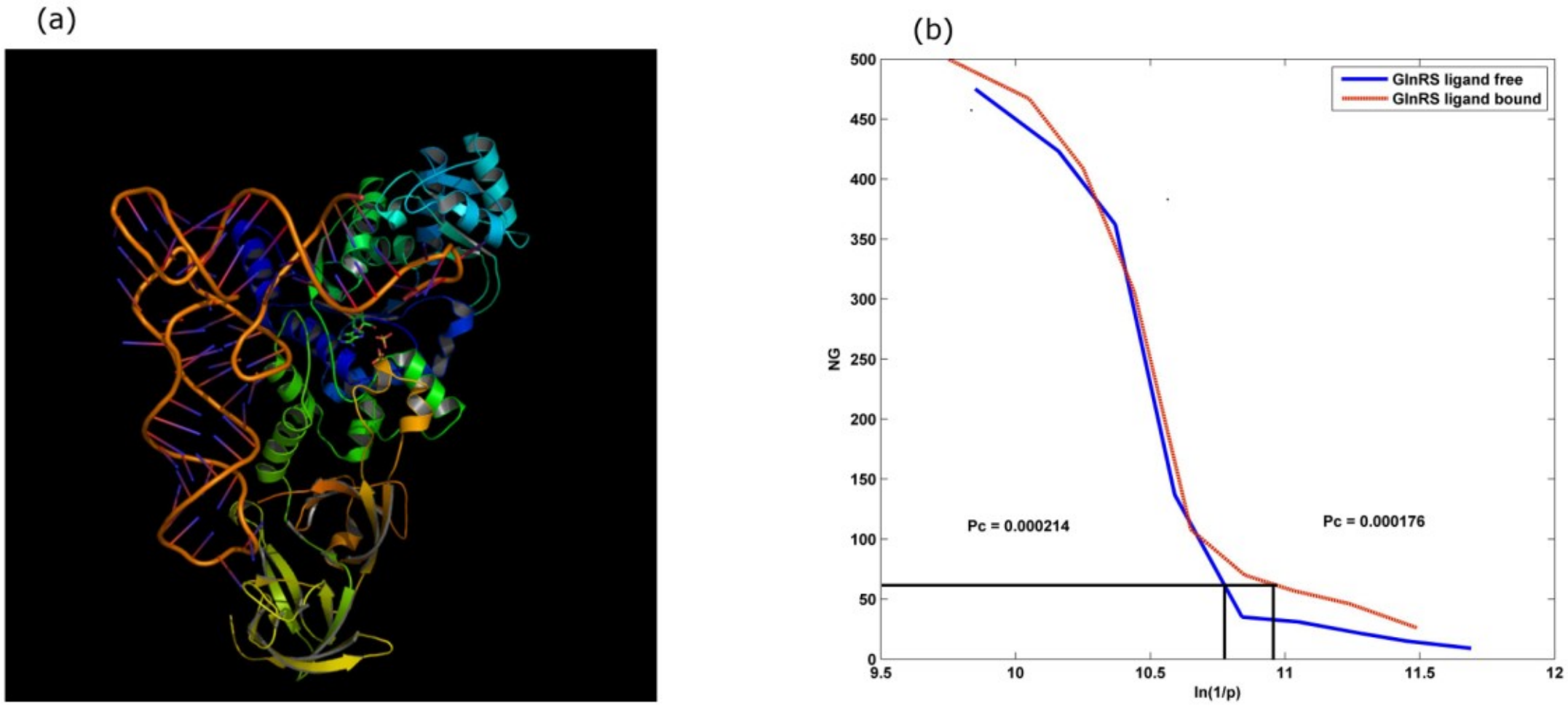

Percolation properties for Glutaminyl t-RNA sythetase in ligand-free form (PDB code: 1NYL) and the fully liganded form (PDB code: 1O0B). (a) The ligand bound protein (in colored cartoon representation) is shown along with the ligands, glutaminyl t-RNA, ATP analogue and Glutamine (in orange cartoon and bonds). The ligand-free form has the same structure but without the tRNA, ATP analog and Glutamine (not shown). (b) The largest cluster ($N_G$) plot versus ln(1/p) for the ligand-free (1NYL) is shown in blue and the ligand-bound form (1O0B) is shown in red. The critical 'p' value ($P_c$) for both cases is indicated in the plot. Although the ligand bound and ligand free forms of the protein are of the same size, they have different $P_c$ values indicating subtle differences in the network properties of the protein influenced by the binding of the ligand.